\documentclass[11pt]{article}
\usepackage{amsmath,color}
\usepackage{amssymb}
\usepackage{amsfonts}
\usepackage{graphicx}
\usepackage{color}

\definecolor{darkgreen}{rgb}{0,0.35,0}

\numberwithin{equation}{section}

\begin{document}

\title{\textbf{Scattering of low lying states in the black hole atmosphere}}
\author{Gaston Giribet}
\date{}
\maketitle

\begin{center}

\smallskip
\smallskip
\centerline{Universit\'{e} Libre de Bruxelles and International Solvay Institutes}
\centerline{{\it ULB-Campus Plaine CPO231, B-1050 Brussels, Belgium.}}

\medskip
\centerline{Departamento de F\'{\i}sica, Universidad de Buenos Aires
FCEN-UBA and IFIBA-CONICET}
\centerline{{\it Ciudad Universitaria, Pabell\'{o}n I, 1428, Buenos Aires, Argentina.}}


\end{center}






\bigskip

\bigskip

\begin{abstract}
We investigate finite $\alpha '$ effects in string theory on a black hole background. By explicitly computing tree level scattering amplitudes, we confirm a duality between seemingly different states recently conjectured by Giveon, Itzhaki, and Kutasov. We verify that the relevant 3-point functions factorize in such a way that the duality between oscillator and winding states becomes manifest. This leads to determine the precise normalization of the dual vertex operators, and confirms at the level of the interacting theory the identification of states suggested by the analysis of the spectrum. This result implies a duality between two seemingly distinct mechanisms driving the violation of the string winding number in the black hole atmosphere. 
\end{abstract}

\newpage

\section{Introduction} 

String theory on a two-dimensional black hole geometry represents an exact solution of the theory \cite{Witten, Mandal} that permits to explore the black hole physics in the deep stringy regime \cite{Stringy, otro, otro2, otro3, otro4, GR, StringyII}; see also \cite{zaka1, zaka2, zaka3, zaka4} and references therein. One of the most interesting properties of this setup is the so-called Fateev-Zamolodchikov-Zamolodchikov (FZZ) duality \cite{FZZ, KKK, HS}, a particular kind of strong/weak duality that associates the string $\sigma $-model on the black hole to a two-dimensional conformal field theory (CFT) consisting of Liouville theory coupled to the sine-Gordon theory, and which consequently receives the name of sine-Liouville theory. Among other results, FZZ duality has been crucial in the formulation of the black hole matrix model \cite{KKK}, which is actually based on the sine-Liouville description. 

From the geometrical point of view, FZZ duality can be thought of as a T-duality that associates the string theory on the Euclidean black hole geometry to the theory formulated on a linear dilaton background and a non-homogeneous tachyon condensate. In terms of the tachyonic picture, several non-perturbative phenomena in the black hole admits a much simpler interpretation; and as it usually happens with strong/weak dualities, it is necessary to consider the two alternative descriptions simultaneously in order to accomplish a fully satisfactory picture.

Recently, the FZZ dualily has been reconsidered and used to investigate the structure of the black hole horizon in string theory \cite{Stringy}, and this led to the discovery of interesting effects: It has been observed in \cite{Stringy} that string states perceive the presence of the horizon in different manners depending on whether or not the momentum in the radial direction is larger than certain critical value set by the string scale $1/\sqrt{\alpha '}$. While at low energy the states experience the horizon in a way that agrees with the semi-classical analysis, at energies sufficiently high the strings start to perceive the horizon as if it had a smeared structure, and this affects the phase-shift in a way that deviates from the behavior of general relativity (GR). The simplest way to see this phenomenon is looking at the reflection coefficient, given by the 2-point function in the black hole geometry. Since in two dimensions the model is exactly solvable, the expression of the 2-point function in known at finite $\alpha '$ \cite{BB}; it exhibits a $\alpha ' $-dependent factor that develops poles at high momentum and produces the anomalous phase-shift. It turns out that such pole conditions, responsible for the stringy behavior, are understood in a much simpler way from the perspective of the FZZ dual picture: In terms of the sine-Liouville model, such poles are seen to come from the integration over the zero-mode of the Liouville direction in the linear dilaton background. This can be rephrased by saying that string states with sufficiently high momentum see the horizon as if it was replaced by the tachyonic profile, what causes the stringy effects and makes them to appear as if they were originated in a region of the moduli space that is behind the horizon \cite{Stringy}. This example shows how the FZZ duality results useful to work out the peculiar details of black hole stringy physics. 

In a recent paper \cite{StringyII}, a generalization of the FZZ correspondence between the black hole geometry and the tachyonic picture has been proposed. This follows from the observation that the string spectrum in the two-dimensional Euclidean black hole background exhibits an intriguing duality between oscillator states and winding string states. As expressed in \cite{StringyII}, this means that normalized states in the black hole geometry have support at widely separated scales, and states that are extended over the black hole atmosphere have a component that is localized near the horizon. While low energy strings probing the black hole see the so-called low lying states as oscillator states in the black hole atmosphere, at large excitation level such states are seen as strings with non-zero winding number around the Euclidean time direction. This relates two seemingly different normalizable modes on the Euclidean background, and it can be seen as a high/low energy correspondence that generalizes FZZ. 

At the level of the spectrum, the duality is related to a particular symmetry of the Hilbert space, the spectral flow symmetry. The string $\sigma $-model on the black hole corresponds to the Wess-Zumino-Witten (WZW) theory for the coset $SL(2,\mathbb{R})/U(1)$, and the spectral flow symmetry is an automorphism of the $\hat{sl}(2)_k$ affine Kac-Moody (KM) algebra that organizes the states in the theory \cite{MO1}. If we denote by $\Phi_{j,m,\bar{m}}$ the vertex operators that create the KM primaries of the $\left\vert  j,m \right\rangle \otimes \left\vert j,\bar{m} \right\rangle $ representation of $SL(2,\mathbb{R})\otimes SL(2,\mathbb{R})$, then the duality between states referred to above translates into the relation between the vertex operators 
\begin{equation}
(J^+)^{\ell }(\bar{J}^+)^{\bar{\ell }}\Phi_{\frac{\ell +\bar{\ell }}{2} -1 ,-\frac{\ell +\bar{\ell }}{2} ,-\frac{\ell +\bar{\ell }}{2} } \label{A11}
\end{equation}
and the vertex operators
\begin{equation}
\Phi_{\frac{k-\ell -\bar{\ell }}{2}-1,\frac{k-\ell +\bar{\ell}}{2},\frac{k+\ell -\bar{\ell}}{2}} \times e^{-\frac{i}{\sqrt{2k}} (({k-\ell +\bar{\ell}}) X-({k+\ell -\bar{\ell}})\bar{X})} \label{A12}
\end{equation}
where $J^+$ in (\ref{A11}) is the current associated to the upper-triangular generators of $SL(2,\mathbb{R})$, and $X$ in (\ref{A12}) is an auxiliary field that represents the $U(1)$ direction of the $SL(2,\mathbb{R})/U(1)$ coset model \cite{BK}. The latter field can be associated to the time compact direction.

The association between operators (\ref{A11}) and (\ref{A12}) is at first sight surprising. While operators (\ref{A11}) represent oscillator states of the gravitational string multiplet, operators (\ref{A12}) correspond to tachyon like string states with both momentum and winding number around the Euclidean time direction (see figure). These two types of states have, in particular, quite different behaviors in the asymptotic far region. This makes the relation between the two pictures particularly interesting.

The idea of the present paper is to test the generalized FZZ duality proposed in \cite{StringyII} at the level of the interacting theory. That is, we are interested in investigating whether the identification between states (\ref{A11}) and (\ref{A12}) persists when interactions are taken into account. The way we will study this is by explicitly computing 3-point string scattering amplitudes in the Euclidean black hole geometry, considering cases in which both operators (\ref{A11}) and (\ref{A12}) are present. This will enable us to verify whether the expressions corresponding to the processes that involve each of the two states actually match. We will begin by briefly reviewing the black hole background and the string spectrum in sections 2 and 3. In section 4, we will compute tree-level string amplitudes involving each of the two operators above and compare the results obtained. Despite the seemingly different interpretation of the oscillator states and winding strings, we will obtain a remarkable matching at the level of 3-point functions. We also discuss the $n$-point functions of operators (\ref{A11}), which are shown to admit an expression in terms of $n$-point correlation functions of Liouville field theory.

\section{Strings on black holes}

The Euclidean black hole solution is given by the following metric and dilaton field
\begin{equation}
ds^2 = 2\ell^2 (dr^2+\tanh^2 r\ d\theta ^2) \ , \ \ \ \ \ \ \ \ e^{-\Phi } =e^{-\Phi_0 } \cosh r , \label{A21}
\end{equation}
where $\ell^2$ is a length scale that controls the size of the asymptotic cylinder to which the geometry tends when the radial direction $r$ is large. Euclidean time $\theta $ has a period that makes the real Euclidean section of the space to be smooth at the horizon, $r=0$. The constant $\Phi_0$ gives the value of the dilaton at the horizon, and it is associated to the mass of the black hole, $M$, as we will discuss below.

The string $\sigma $-model on the background (\ref{A21}) corresponds to the WZW model for the coset $SL(2,\mathbb{R})/U(1)$ \cite{Witten}, and thus it enjoys KM affine symmetry. The central charge of this model is
\begin{equation}
c=2+\frac{6}{k-2},
\end{equation}
where $k=\ell^2/\alpha'$ is the level of the WZW action and thus controls the quantum effects in the worldsheet theory. Large $k$ describes a weakly curved black hole, and one expects the semi-classical intuition to be valid in that limit.
\begin{figure}
\ \ \ \ \ \ \ \ \ \  \ \ \ \ \ \ \includegraphics[width=4in]{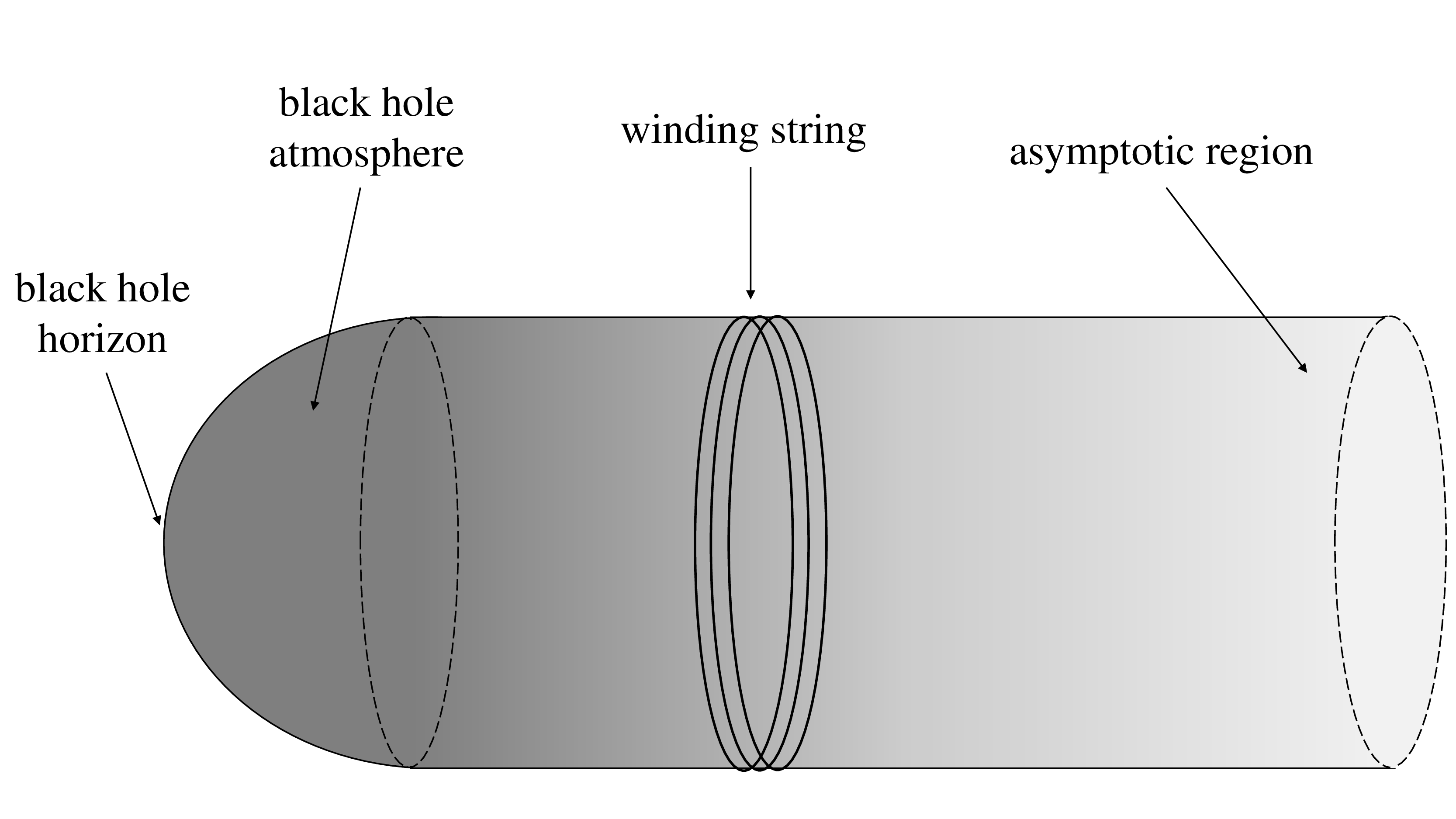}
\caption{The Euclidean black hole geometry resembles a semi-infinite cigar whose compact direction represents the periodic Euclidean time. The black hole horizon is located at the tip of the cigar ($r=0$). The black hole atmosphere corresponds to the cap, a region where the curvature in string units is of order $1/k$. Asymptotically, the geometry tends to a cylinder of radius $\sqrt{k}$ in string units. Strings can have non-zero winding number around the time direction $\theta $. Such winding number, however, can be violated in a scattering process due to the topology of the space.}
\end{figure}

\section{String spectrum}

The spectrum of string states in the Euclidean black hole geometry (\ref{A21}) is parameterized by indices $j,$ $m,$ $\bar{m}$ that labels the vectors $\left\vert  j,m\right\rangle \otimes \left\vert  j,\bar{m}\right\rangle $ that form representations of $SL(2,\mathbb{R})\otimes {SL}(2,\mathbb{R})$. More precisely, one has that the momentum of the states along the compact direction $\theta $ and the winding number around that direction are given by 
\begin{equation}
p = \bar{m} - m  \ , \ \ \ \ \ \ \  
\omega  = \frac{\bar{m} + m}{k},
\end{equation}
respectively; the momentum in the radial direction is given by $j$. The winding number $\omega $ is a conserved quantity only at large $r$, while it can be violated for states that probe the tip of the Euclidean geometry (see figure). 

The representations of $\hat{sl}(2)_k$ affine algebra that give the Hilbert space of the theory are defined by how the KM modes $J_n^a$, $\bar{J}_n^a$ (with $a={\pm, 3}$) act on the vectors; namely 
\begin{eqnarray}
J_0^{\pm }\left|j,m\right\rangle &=& (\pm j\pm 1 -m) \left|j,m\pm 1 \right\rangle ,  \ \ \ \ \ \ \ \ J_{n>0}^{\pm }\left|j,m\right\rangle = 0 , \label{A32} \\
J_0^{3 }\left|j,m\right\rangle &=& m \left|j,m \right\rangle ,  \ \ \ \ \ \ \ \  \ \ \ \ \ \ \ \ \ \ \ \  \ \ \ \  \ \ \ \  J_{n>0}^{3 }\left|j,m\right\rangle = 0 , \label{A33} 
\end{eqnarray}
together with the anti-holomorphic (bared) counterpars. This defines highest-weight KM representations. Unitarity and redundancy due to the spectral flow isomorphism demand the constraint $-1\leq 2j \leq k-3$, with $m-j$ and $\bar{m}-j$ being positive integers. In addition, the construction of the coset theory requires $J^3$ and $\bar{J}^3$ to vanish on the physical states. We will see below how to implement this constraint in an efficient way.

Conditions (\ref{A32})-(\ref{A33}) can be rephrased in terms of the operator product expansion (OPE) between the local currents 
\begin{equation}
J^{a}(z)=\sum_{n\in \mathbb{Z}}J_{n}^{a}\ z^{-1-n}\ , 
\end{equation}
and the vertex operators $\Phi _{j,m,\bar{m}} (z)$ that create the states $\left\vert  j,m\right\rangle \otimes \left\vert  j,\bar{m}\right\rangle $ from the vacuum of excitations. That is, 
\begin{eqnarray}
J^{\pm }(z)\Phi_{j,m,\bar{m}} (w) &=& \frac{(\pm j\pm 1 -m)}{(z-w)} \Phi_{j,m\pm 1,\bar{m}} (w)+ ...  \ , \\ 
J^{3}(z)\Phi_{j,m,\bar{m}} (w) &=& \frac{m}{(z-w)} \Phi_{j,m,\bar{m}} (w)+ ...  \label{OPE}
\end{eqnarray}
and the anti-holomorphic counterpart; here, the ellipses stand for regular terms.  

A convenient realization of this algebra is given by the Wakimoto free field representation \cite{Wakimoto}, which suffices to realize the KM currents ${J}^{a}$, $\bar{J}^{a}$ in terms of a free scalar field $\phi $ with background charge, and a $(\beta , \gamma )$ ghost system; namely
\begin{equation}
J^+=\beta \ , \ \ \ \ J^-= \beta \gamma^2+\frac{2}{Q} \gamma \partial \phi +k\partial \gamma   \ , \ \ \ \ J^3=-\beta \gamma -\frac{1}{Q} \partial \phi \label{J}
\end{equation}
with the free field correlators $\left\langle \phi (z)\phi (w)\right\rangle =-\log (z-w)$, $\left\langle \beta (z)\gamma (w)\right\rangle =1/(z-w)$, and with the background charge $Q$ given by 
\begin{equation}
Q=\sqrt{\frac{2}{k-2}} .
\end{equation}

Currents (\ref{J}) can be shown to realize the $\hat{sl}(2)_k$ affine algebra. The vertex operators $\Phi _{j,m,\bar{m}}$ in terms of the Wakimoto representation take the form 
\begin{equation}
\Phi_{j,m,\bar{m}} = \gamma ^{-1-j-m} \bar\gamma ^{-1-j-\bar{m}}e^{-Q(j+1)\phi } \label{A39}
\end{equation}
which can be seen to coincide with the large $\phi $ behavior of normalizable fields in the $SL(2,\mathbb{R})$ WZW model. These operators realize the OPE (\ref{OPE}).

In terms of Wakimoto free fields, the action of the $SL(2,\mathbb{R})$ WZW model reads
\begin{equation}
S_{SL(2,\mathbb{R})}=\frac{1}{4\pi }\int d^2z \Bigg( \partial \phi \bar{\partial }\phi -QR\phi +\beta \bar{\partial}\gamma +\bar{\beta }\partial \bar{\gamma} -4\pi M \beta \bar{\beta } e^{-Q\phi}\Bigg). \label{action}
\end{equation}

The last term in the Lagrangian is responsible for the interaction, which becomes weak in the large $\phi $ region. In these variables, large $\phi $ corresponds to the region far from the black hole horizon. In this spacetime interpretation of the coset, the constant $M$ is associated to the black hole mass. Its value can be set to 1 by shifting the zero-mode of $\phi $. This is why we said before that $M$ is related to the value of the dilaton at the horizon, namely $\delta \log M=-\delta \Phi$. 

As mentioned above, in order to construct the theory on the coset one has to impose that both ${J}_0^{3}$ and $\bar{J}_0^{3}$ annihilate the physical states. A practical way of implementing this condition is to consider an extra free scalar field $X$ and a $(B,C)$ ghost system \cite{BK, DVV, BB, GN2}. This allows to define the BRST charge operator
\begin{equation}
Q_{U(1)} = \oint dz\ C(J^3+i\sqrt{k/2}\partial X),
\end{equation}
and define the vertex operators for the theory on the coset by dressing the operators $\Phi_{j,m,\bar{m}}$ as follows
\begin{equation}
V_{j,m,\bar{m}} = \Phi_{j,m,\bar{m}} \times e^{-i\sqrt{\frac{2}k}(mX -\bar{m} \bar{X})} . \label{V}
\end{equation}

This operators have vanishing $U(1)$ charge and have the right conformal dimension for the coset theory, namely 
\begin{equation}
\Delta = -\frac{j(j+1)}{k-2}+\frac{m^2}{k} \ , \ \ \ \ \ \bar{\Delta } = -\frac{j(j+1)}{k-2}+\frac{\bar{m}^2}{k}\  .
\end{equation}

Other operators of the coset theory that play a crucial role in the discussion are
\begin{equation}
\tilde{V}_{\ell , \bar{\ell }}=(J^+)^{\ell }(\bar{J}^+)^{\bar{\ell }}\Phi_{\frac{\ell +\bar{\ell }}{2} -1 ,-\frac{\ell +\bar{\ell }}{2} ,-\frac{\ell +\bar{\ell }}{2} } . \label{O}
\end{equation}

These are the oscillator operators mentioned in the introduction. These have conformal dimension 
\begin{equation}
\Delta = \ell-\frac{(\ell +\bar{\ell})(\ell +\bar{\ell}-2)}{4(k-2)} \ , \ \ \ \ \ \bar\Delta = \bar{\ell }-\frac{(\ell +\bar{\ell})(\ell +\bar{\ell}-2)}{4(k-2)},
\end{equation}
which coincides with the conformal dimension of a state of momenta $j=(k-\ell -\bar{\ell })/2-1$, $m=(k-\ell +\bar{\ell})/2$, $\bar{m}=(k+\ell -\bar{\ell})/2$; in other words, (\ref{O}) has the same observables that a state with winding number $\omega = 1$ and momentum $p=\ell -\bar{\ell }$ around the $\theta $ direction. This is the origin of the identification between states proposed in \cite{StringyII}. At the level of the spectrum, the operator (\ref{O}) and the operator $V_{({k-\ell -\bar{\ell }})/{2}-1,({k-\ell +\bar{\ell}})/{2},({k+\ell -\bar{\ell}})/{2}}$ exhibit similar properties despite the fact that the spacetime interpretation of each of them is quite different.

Operator (\ref{O}) also admits a simple representation in terms of the Wakimoto free fields; namely
\begin{equation}
\tilde{V}_{\ell , \bar{\ell }}= N_{\ell} N_{\bar{\ell }} \ \beta^{\ell }\bar{\beta }^{\bar{\ell }} e^{-\frac{Q}{2}(\ell + \bar{\ell })\phi } \label{A316}
\end{equation}
where $N_{\ell }$ is a normalization factor. This follows from (\ref{J}). This type of operator was originally considered in \cite{BK} in the context of the stringy black hole. More recently, a similar representation has been considered in relation to string theory on AdS$_3$ space \cite{GiribetLopez, Gaston2, Gaston3}. According to \cite{StringyII}, operator (\ref{O}) would be dual to the tachyon like operator
\begin{equation}
V_{\frac{k-\ell -\bar{\ell }}{2}-1,\frac{k-\ell +\bar{\ell}}{2},\frac{k+\ell -\bar{\ell}}{2}} = \gamma^{\ell -k }\bar{\gamma}^{\bar{\ell }-k} e^{{Q}(\frac{\ell+\bar{\ell }-k}{2} )\phi -i\sqrt{\frac{k}{2}} (X-\bar{X})
+i\frac{(\ell - \bar{\ell } )}{\sqrt{2k}} (X+\bar{X})}, \label{A317}
\end{equation}
which represents a winding state. Notice that, in particular, these two operators exhibit totally different behavior at large $\phi $. Aimed at verifying whether the identification between oscillator and winding states discussed in \cite{StringyII} also holds when one includes interactions, here we will calculate string amplitudes that involve both operators (\ref{A11}) and (\ref{A12}). Having the explicit representation (\ref{A316})-(\ref{A317}) at hand, we are in principle able to compute such amplitudes explicitly. 

\section{String amplitudes} 

Let us begin by computing string amplitudes that involve states associated to operators (\ref{A316}). One such an amplitude is given by
\begin{equation}
{\mathcal A}_{\ell_1 , j_2, j_3, ... , j_n} = \int \prod_{j=1}^{n} \frac{d^2z_j}{\text{Vol}_{{PSL}(2,\mathbb{C})}} \Bigg\langle :(J^+)^{\ell_1}(\bar{J}^+)^{\ell_1}\Phi_{\ell_1-1 ,\ell_1 ,\ell_1 }(z_1): \  \prod_{i=2}^n :V_{j_i,m_i,\bar{m}_i}(z_i):  \Bigg\rangle \label{first}
\end{equation}
where the correlation function is defined for the theory (\ref{action}) on the Riemann sphere. However, this is not exactly the correlator we have to take a look at. Correlator (\ref{first}) describes a process in which $\sum_{i=2}^n m_i=\sum_{i=2}^n \bar{m}_i=0$, while what we are interested in here is a process involving $\tilde{V}_{\ell_1 , {\ell }_1}$ in which the conservation law is rather $\sum_{i=2}^n m_i=\sum_{i=2}^n \bar{m}_i=-k/2$, meaning $1+\sum_{i=2}^n \omega_i =0$. This is because the comparative analysis we want to perform is between a correlator involving $\tilde{V}_{\ell_1 , \bar{\ell }_1}$ and a correlator involving the state $V_{k/2-1-\ell_1,k/2,k/2}$, and the latter has winding number $\omega_1 =1$. Therefore, we have to modify the correlator $\tilde{\mathcal A}_{\ell_1 , ... , j_n}$ above in such a way that it allows for the violation of the total winding number in one unit. As already mentioned, in this scenario the violation of the total winding number $ \sum_{i=1}^n\omega_i$ does not represent any conceptual problem, as it can be understood in a simple way due to the contractibility of the Euclidean black hole geometry. However, from the technical point of view, computing winding violating correlators is non-trivial. There is, however, a very interesting prescription for defining such correlators proposed by Fateev and the brothers Zamolodchikov in an unpublished paper \cite{FZZ} (see also \cite{MO3}). The method proposed by FZZ amounts to introduce in the 3-point correlation function a fourth operator of conformal dimension zero. Such operator is interpreted as a {\it conjugate identity} operator, and in the Wakimoto free field representation it reads
\begin{equation}
V_{-\frac{k}{2},\frac{k}{2},\frac{k}{2}} = \gamma^{-1}\bar{\gamma}^{-1}e^{\frac{\phi}{Q}-i\sqrt{\frac{k}{2}}(X-\bar{X})} . \label{uno}
\end{equation}
One can easily verify that this operator satisfies $\Delta=\bar{\Delta}=0$ and is a good operator on the coset. The recipe \cite{FZZ} is to introduce (\ref{uno}) inside the correlator and then remove the factor that depends on its inserting point $z_4$. That is to say, the relevant 3-point function we are interested in is
\begin{equation}
\tilde{\mathcal A}_{\ell_1 , j_2 , j_3} =   \Bigg\langle \tilde{V}_{\ell_1 , \ell_1}(0) V_{j_2,m_2,\bar{m}_2}(1) V_{-\frac{k}{2},\frac{k}{2},\frac{k}{2}}(z_4) V_{j_3,m_3,\bar{m}_3}(\infty ) \Bigg\rangle , \label{posta}
\end{equation}
which now satisfies the desired charge condition $m_2+m_3=\bar{m}_2+\bar{m}_3=-k/2$. Here, we have set $z_1=0$, $z_2=1$, $z_3 = \infty$ in order to cancel the volume of the conformal Killing group. Hereafter, we omit the symbols :: of normal ordering.

A notable simplification follows from choosing in (\ref{posta}) the kinematical configuration $m_2=\bar{m}_2=-1-j_2$. This renders the combinatorics of the Wick contraction of $(\beta, \gamma)$ fields much simpler without representing a severe loss of generality.

The particular correlators (\ref{posta}) has not been computed in the literature. However, its calculation is accessible with the techniques developed in \cite{BB, GN2, GN3, GiribetLopez, Gaston2, Gaston3}, which amounts to perform a Coulomb gas computation using Wakimoto representation (\ref{A39}) and (\ref{A316}). It is a common misunderstanding to think that such a computation would not yield the exact result but merely an approximation valid at large $\phi $. In fact, it has been shown in reference \cite{GN3} that this approach actually reproduces the exact expressions \cite{Teschner}. The reason why the Coulomb gas computation suffices to give the exact result is that, despite the fact that treating the interaction term in the action as a perturbation is only valid at large $\phi $, it turns out that the integration over the zero mode of $\phi $ in the path integral calculation yields a $\delta $ function that singles out only one term in the expansion of the exponentiation of the interaction. This implies that no large $\phi $ approximation is actually needed. As a result, the problem of solving the correlator reduces to that of performing a multiple integrals of the Fateev-Dotsenko type \cite{DF} and, after that, analytically continuing the result to generic values of the external momenta. Such analytic continuation is under control and it has been successfully carried out in the literature in diverse examples. In the case of correlator (\ref{posta}), the integral representation reads
\begin{eqnarray}
{\mathcal A}_{\ell_1 , j_2 , j_3}  &=& \Gamma (-s) M^{s}  N^2_{\ell_1 }{\Gamma^2 (-j_3-m_3)} \  {\mathcal I}_s(\ell , j_2)
\end{eqnarray}
with
\begin{eqnarray}
{\mathcal I}_s(\ell_1 , j_2)=\int \prod_{r=1}^{s} d^2w_r \prod_{t=1}^s \Bigg( |w_t|^{-\frac{4}{k-2}\ell_1}  |1-w_t|^{-\frac{4}{k-2}(j_2+1)} \prod_{r=1}^{t-1} |w_r - w_t |^{-\frac{4}{k-2}} \Bigg) \label{Isl}
\end{eqnarray}

In this expression, the amount of integrals to be performed is given by $s=k/2-2-\ell_1-j_2-j_3$. This means that, as it is written in (\ref{Isl}), the expression only make sense for $\ell_1+j_2+j_3-k/2\in \mathbb{Z}_{< -1}$. However, this restriction can be easily circumvented as we discuss below.

A remarkable property of operator (\ref{uno}) is that its OPE with the interaction term in the action does not produce singularities at $z_4\to w_r$. This is because the contribution of the field $\phi $ and of the ghost system to the poles mutually cancel. More precisely, we have
\begin{equation}
V_{\varepsilon-\frac{k}{2} ,\frac{k}{2} ,\frac{k}{2} } (z_4) \tilde{V}_{1,1} (w_r) \simeq |z_4-w_r|^{-\frac{4\varepsilon }{k-2}} 
\end{equation}
The relevant effect produced by the presence of operator $V_{-k/2,k/2,k/2}$ is that of affecting the integration over the zero-mode of the fields in such a way that the charge compensation condition becomes $1+\omega_2+\omega_3=0$, as desired.

It turns out that the multiple integral (\ref{Isl}) can be solved explicitly \cite{DF}. It yields
\begin{eqnarray}
\tilde{\mathcal A}_{\ell_1, j_2, j_3} &=& \Gamma(-s) \Gamma (s+1)  {\Gamma^2 (-j_3-m_3)} N^2_{\ell_1 } \pi^s M^s \frac{\Gamma^s(1+\frac{1}{k-2})}{\Gamma^s(-\frac{1}{k-2})}  \times  \nonumber \\ 
&&\times \prod_{t=1}^{s} \Bigg(
\frac{\Gamma (-\frac{t}{k-2})}{\Gamma (1+\frac{t}{k-2})}
\frac{\Gamma (-\frac{2j_3+1+t}{k-2}) \Gamma (1-\frac{2\ell_1-1+t}{k-2}) \Gamma (1-\frac{2j_2+1+t}{k-2})}{\Gamma (1+\frac{2j_3+1+t}{k-2}) \Gamma (\frac{2\ell_1-1+t}{k-2}) \Gamma (\frac{2j_2+1+t}{k-2})}\Bigg)
\end{eqnarray}

As for (\ref{Isl}), this expression only makes sense for values such that $s\in \mathbb{Z}_{>0}$. Amplitudes that correspond to other values of $s$ then require analytic continuation. In order to perform such continuation, it is convenient to write the result in terms of the special function $\Upsilon $, introduced by the brothers Zamolodchikov in reference \cite{ZZ} in the context of Liouville field theory. Defining $\hat{\Upsilon }(x)\equiv {\Upsilon }(bx)$ with $b^{-2}=k-2$, the final expression for the correlator above reads
\begin{eqnarray}
\tilde{\mathcal A}_{\ell_1,j_2,j_3} &=& 
- N^2_{\ell_1 }{\Gamma^2 (-j_3-m_3)}  \Bigg( - \pi M (k-2)^{\frac{1}{k-2}} \frac{\Gamma (1+\frac{1}{k-2})}{\Gamma (1-\frac{1}{k-2})}   \Bigg)^{{k}/{2}-2-\ell_1 -j_2 -j_3 } \times \nonumber \\ 
&& \times \ \frac{\Upsilon' (0) \hat{\Upsilon } (2\ell _1)  }{\hat{\Upsilon } (\ell_1+j_2+j_3-k/2+2) \hat{\Upsilon } (-\ell_1 +j_2+j_3+k/2+1) } \times \nonumber \\ 
&& \times \ \frac{ \hat{\Upsilon } (2j_2+2) \hat{\Upsilon } (2j_3+k) }{\hat{\Upsilon } (\ell_1-j_2+j_3+k/2-1) \hat{\Upsilon } (\ell_1 +j_2-j_3-k/2+1)} 
 \label{firsto}
\end{eqnarray}
where $\Upsilon '(x)=\frac{d}{dx}\Upsilon(x)$. Function $\hat{\Upsilon }(x)$ has its zeroes at $x=-m(k-2)-n$ and $x=(m+1)(k-2)+(n+1)$ with $m,n\in \mathbb{Z}_{> 0}$.

The next step in our analysis is to compute a second correlator; namely, the correlator that is dual to (\ref{posta}) in the sense of being defined by replacing the vertex $\tilde{V}_{\ell_1 , \ell_1}$ by the vertex $V_{k/2-1-\ell_1,k/2,k/2}$. That is, we want to compute
\begin{eqnarray}
{\mathcal A}_{\ell_1,j_2,j_3} = \Bigg\langle V_{\frac{k}{2}-1-\ell_1,\frac{k}{2} ,\frac{k}{2} }(0) V_{j_2,-1-j_2,-1-j_2}(1) V_{j_3,m_3,\bar{m}_3}(\infty ) \Bigg\rangle \label{elotro}
\end{eqnarray}
This can be done in a similar way as before. The integral representation obtained in this case, however, presents several difference with respect to that of $\tilde{\mathcal A}_{\ell_1 , j_2 , j_3}$. The main differences are three: First, the factor in $|w_r |^{-\frac{4}{k-2}\ell_1}$ in the integrand (\ref{Isl}) now changes to $|w_r |^{\frac{4}{k-2}(\ell_1-{k}/{2})-2}$, with the last $-2$ in the new exponent coming from the contraction of the fields $\gamma $ in (\ref{A317}), which are absent in the case of (\ref{A316}). Secondly, the amount of integrals to be performed in the case of ${\mathcal A}_{\ell_1 ,j_2 ,j_3}$ is not $s$, but $s+2\ell_1 -k+1$. In the third place, the $m$-dependent multiplicity factor coming from the combinatorics of the Wick contraction of the $(\beta , \gamma )$ fields also changes. 

Remarkably, despite all these differences, after a relatively lengthy computation one obtains
\begin{eqnarray}
{\mathcal A}_{\ell_1,j_2,j_3} &=& 
 \frac{\Gamma (-j_3-m_3)\Gamma(\ell_1-k+1)}{\Gamma(1+j_2+m_3)\Gamma (k-\ell_1)  }  \Bigg(  \pi M (k-2)^{\frac{1}{k-2}} \frac{\Gamma (1+\frac{1}{k-2})}{\Gamma (1-\frac{1}{k-2})}   \Bigg)^{-{k}/{2}-1+\ell_1 -j_2 -j_3 } \times  \nonumber \\
&& \times \ \frac{\Upsilon' (0) \hat{\Upsilon } (2\ell _1-k+1) }{\hat{\Upsilon } (\ell_1+j_2+j_3-k/2+2) \hat{\Upsilon } (-\ell_1 +j_2+j_3+k/2+1) } \times  \nonumber \\
&& \times \
\frac{\hat{\Upsilon } (2j_2+2) \hat{\Upsilon } (2j_3+k) }{\hat{\Upsilon } (\ell_1-j_2+j_3+k/2-1) \hat{\Upsilon } (\ell_1 +j_2-j_3-k/2+1)}
,
\end{eqnarray}
which actually looks pretty similar to the result (\ref{firsto}), at least in what regards the dependences in the $\hat{\Upsilon}$ functions.

In order to further simplify the expressions, one may set the value of the black hole mass as follows
\begin{equation}
M=\frac{1}{\pi (k-2)} \frac{\Gamma (-\frac{1}{k-2})}{\Gamma (\frac{1}{k-2})},
\end{equation}
which is easily achieved by shifting the zero mode of $\phi $. We already mentioned that the value of the dilaton at the horizon controls the black hole mass.

Now, with the final expressions of both correlators at hand, it only remains to make the comparative analysis. Taking into account that $\Gamma (1+z-n) = (-1)^n {\Gamma (1+z) \Gamma (-z)}/{\Gamma (n-z)}$ for $z\in \mathbb{C}$, $n\in \mathbb{Z}_{\geq 0}$, and using the properties of $\Upsilon$ function under shift of its argument in $b^{\pm 1}$ units, one arrives to the formula
\begin{equation}
\tilde{\mathcal A}_{\ell_1,j_2,j_3}= \frac{N_{\ell_1}^2 (k-2)^2 \Gamma^{2}(k-\ell_1)}{(2\ell_1-1)}\frac{ 
\Gamma (\frac{2\ell_1-1}{k-2}) \Gamma (2\ell_1-k+1)}{
\Gamma (-\frac{2\ell_1-1}{k-2}) \Gamma (k-2\ell_1)
}  \ {\mathcal A}_{\ell_1,j_2,j_3}   . \label{formulafinal}
\end{equation}

This means that, remarkably enough, the quotient $\tilde{\mathcal A}_{\ell_1 , j_2 ,j_3} / {\mathcal A}_{\ell_1 , j_2 ,j_3}$ happens to depend only on the momentum $\ell _1$, whose dependence can be absorbed in the normalization of the vertices. All dependences of the other momenta cancel out. This factorization implies, in particular, that the pole structure that allows to extract physical information from the correlators, such as the fusion rules, is the same in both cases. This manifestly shows that the correspondence between states (\ref{A11}) and (\ref{A12}) gets realized at the level of string amplitudes.

Let us notice that the relative factor in (\ref{formulafinal}) is reminiscent of the reflection coefficient of Liouville field theory, usually denoted by $R(\alpha )$, provided one identifies $\alpha = b\ell_1$ and $b=Q/\sqrt{2}$. We can see that this is actually not an accident: So far, we have been concerned with $3$-point functions. This is because $3$-point functions are the simplest non-trivial observables that one is able to compute using (\ref{Isl}) and express explicitly in terms of known functions.  
However, there are interesting properties of the $n$-point functions that can be read from the integral representation even when a closed expression in terms of special functions is not available for $n>3$. One such a property is the fact that the $n$-point functions involving operators (\ref{A316}) in the coset theory admit to be expressed in terms of the $n$-point function in Liouville field theory (LFT) in a remarkably simple manner. The precise relation is given by the formula
\begin{equation}
\Bigg\langle \prod_{i=1}^{n-1}\tilde{V}_{\ell_i, \ell_i}(z_i) \ V_{\ell_n , 0 , 0}(\infty )\Bigg\rangle_{\frac{SL(2,\mathbb{R})}{U(1)}} = \ \prod_{j=1}^{n}\Gamma^{2}(\ell_j+1) \ \Bigg\langle \prod_{i=1}^{n-1}{V}^L_{\alpha_i}(z_i) \ V^L_{Q_L-\alpha_n }(\infty )\Bigg\rangle_{\text{LFT}}  \label{chorongo}
\end{equation}
with the dictionary
\begin{equation}
Q_L=b+\frac{1}{b} , \ \ \ \ \ \ \ \ b^{2}= \frac{1}{k-2} \ , \ \ \ \ \ \ \ \ \alpha_j = b\ell_j , \ \ \ \ \ \ \ \ N_{\ell_j } = {\Gamma(\ell_j +1)} , \label{chorizo}
\end{equation}
where $V^{L}_{\alpha }=e^{\sqrt{2}\alpha \varphi }$ are the exponential primary operators of Liouville field theory, the latter being defined by the Liouville action
\begin{equation}
S_{L}=\frac{1}{4\pi }\int d^2z \Bigg( \partial \varphi \bar{\partial }\varphi +Q_LR\varphi -4\pi  e^{\sqrt{2}b\varphi}\Bigg).
\end{equation}
From (\ref{chorizo}), one verifies that the conformal dimension of the fields in the correlator on the left hand side coincides with the conformal dimension of Liouville theory, $\Delta =\alpha (Q_L - \alpha )$. Relation (\ref{chorongo}) is reminiscent of the $H_3^+$-Liouville correspondence \cite{RT}. A relation closely related to (\ref{chorongo}) has been discussed in the context of string theory on AdS$_3$ space in \cite{Gaston2, Gaston3}. This provides a tool to study properties of string amplitudes in the Euclidean black hole geometry in terms of the much better understood Liouville CFT. In the case of the 3-point amplitudes, here this led us to provide evidence for the identification of states (\ref{A11}) and (\ref{A12}) at the level of the interacting theory.

\[ \]
This work has been supported by CONICET through the grant PIP 0595/13. The author thanks the hospitality of Pontificia Universidad Cat\'olica de Valpara\'{\i}so (PUCV) for the hospitality during his stay, during which this work has been completed.
\[ \]

\providecommand{\href}[2]{#2}\begingroup\raggedright\endgroup

\begin{thebibliography}{10}


\bibitem{Witten}
  E.~Witten,
  ``On string theory and black holes,''
  Phys.\ Rev.\ D {\bf 44}, 314 (1991).
	
\bibitem{Mandal}
  G.~Mandal, A.~M.~Sengupta and S.~R.~Wadia,
  ``Classical solutions of two-dimensional string theory,''
  Mod.\ Phys.\ Lett.\ A {\bf 6}, 1685 (1991).

\bibitem{Stringy} 
  A.~Giveon, N.~Itzhaki and D.~Kutasov,
  ``Stringy Horizons,''
  JHEP {\bf 1506}, 064 (2015).

\bibitem{otro} 
  R.~Ben-Israel, A.~Giveon, N.~Itzhaki and L.~Liram,
  ``Stringy Horizons and UV/IR Mixing,''
  JHEP {\bf 1511}, 164 (2015).

\bibitem{otro2} 
  R.~Ben-Israel, A.~Giveon, N.~Itzhaki and L.~Liram,
  ``On the Stringy Hartle-Hawking State,''
  JHEP {\bf 1603}, 019 (2016).

\bibitem{otro3} 
  A.~Giveon and N.~Itzhaki,
  ``String Theory Versus Black Hole Complementarity,''
  JHEP {\bf 1212}, 094 (2012).

\bibitem{otro4} 
  A.~Giveon and N.~Itzhaki,
  ``String theory at the tip of the cigar,''
  JHEP {\bf 1309}, 079 (2013).

\bibitem{GR} 
  G.~Giribet and A.~Ranjbar,
  ``Screening Stringy Horizons,''
  Eur.\ Phys.\ J.\ C {\bf 75}, no. 10, 490 (2015).
	
\bibitem{StringyII} 
  A.~Giveon, N.~Itzhaki and D.~Kutasov,
  ``Stringy Horizons II,''
  arXiv:1603.05822 [hep-th].

\bibitem{zaka1}
  T.~G.~Mertens, H.~Verschelde and V.~I.~Zakharov,
  ``Random Walks in Rindler Spacetime and String Theory at the Tip of the Cigar,''
  JHEP {\bf 1403}, 086 (2014).

\bibitem{zaka2}
  T.~G.~Mertens, H.~Verschelde and V.~I.~Zakharov,
  ``Perturbative String Thermodynamics near Black Hole Horizons,''
  JHEP {\bf 1506}, 167 (2015).

\bibitem{zaka3}
  T.~G.~Mertens, H.~Verschelde and V.~I.~Zakharov,
  ``The long string at the stretched horizon and the entropy of large non-extremal black holes,''
  JHEP {\bf 1602}, 041 (2016).
  
\bibitem{zaka4}
  T.~G.~Mertens, H.~Verschelde and V.~I.~Zakharov,
  ``Hagedorn temperature and physics of black holes,''
  arXiv:1605.02785 [hep-th].
  
\bibitem{FZZ} V. Fateev, A. Zamolodchikov, and Al. Zamolodchikov,
unpublished.

\bibitem{KKK} 
  V.~Kazakov, I.~K.~Kostov and D.~Kutasov,
  ``A Matrix model for the two-dimensional black hole,''
  Nucl.\ Phys.\ B {\bf 622}, 141 (2002)
	
\bibitem{HS} 
  Y.~Hikida and V.~Schomerus,
  ``The FZZ-Duality Conjecture: A Proof,''
  JHEP {\bf 0903}, 095 (2009).	
		
\bibitem{BB} K. Becker and M. Becker, ``{Interactions in the
SL(2,R)/U(1) Black Hole Background},'' Nucl. Phys. B \textbf{418}, 206 (1994).

\bibitem{MO1} J.M. Maldacena and H. Ooguri, ``Strings in AdS$_3$ and the SL(2,R) WZW Model. Part 1: The Spectrum,'' J.
Math. Phys. \textbf{42}, 2929 (2001).

\bibitem{BK} M. Bershadsky and D. Kutasov, ``{Comment of gauged WZW
theory},'' Phys. Lett. B \textbf{266}, 345 (1991).

\bibitem{Wakimoto} M. Wakimoto, ``{Fock representations of the affine
Lie algebra A(1)}$^{1}$,'' Comm. Math. Phys. \textbf{111}, 75 (1986).

\bibitem{DVV}
  R.~Dijkgraaf, H.~L.~Verlinde and E.~P.~Verlinde,
  ``String propagation in a black hole geometry,''
  Nucl.\ Phys.\ B {\bf 371}, 269 (1992).
	
\bibitem{GiribetLopez} G. Giribet and D. L\'{o}pez-Fogliani, ``{Remarks
on free field realization of SL(2,R)/U(1) }$\mathit{\times }${\ U(1)
WZNW model},'' JHEP \textbf{0406}, 026 (2004).

\bibitem{Gaston2} G. Giribet, ``{Violating the string winding number
maximally in Anti-de Sitter space},'' Phys. Rev. D \textbf{84}, 024045 (2011).

\bibitem{Gaston3} 
  G.~Giribet,
  ``One-loop amplitudes of winding strings in AdS$_3$ and the Coulomb gas approach,''
  Phys.\ Rev.\ D {\bf 93}, no. 6, 064037 (2016).
	
\bibitem{MO3} J.M. Maldacena and H. Ooguri, ``Strings in AdS$_{3}$ and the SL(2,R) WZW Model. Part 3: Correlation
Functions,'' Phys. Rev. D \textbf{65}, 106006 (2002).


\bibitem{GN2} G. Giribet and C. N\'{u}\~{n}ez, ``{Aspects of the free
field description of string theory on AdS}$_{3}$,'' JHEP \textbf{0006}, 033
(2000).

\bibitem{GN3} G. Giribet and C. N\'{u}\~{n}ez, ``{Correlators in AdS}$_{3}${\ string theory},'' JHEP \textbf{0106}, 010 (2001).

\bibitem{Teschner} 
  J.~Teschner,
  ``On structure constants and fusion rules in the SL(2,C) / SU(2) WZNW model,''
  Nucl.\ Phys.\ B {\bf 546}, 390 (1999).

\bibitem{DF}
  V.~S.~Dotsenko and V.~A.~Fateev,
  ``Four Point Correlation Functions and the Operator Algebra in the Two-Dimensional Conformal Invariant Theories with the Central Charge $c<1$,''
  Nucl.\ Phys.\ B {\bf 251}, 691 (1985).

\bibitem{ZZ} 
  A.~B.~Zamolodchikov and A.~B.~Zamolodchikov,
  ``Structure constants and conformal bootstrap in Liouville field theory,''
  Nucl.\ Phys.\ B {\bf 477}, 577 (1996).

\bibitem{RT}	
  S.~Ribault and J.~Teschner,
  ``H+(3)-WZNW correlators from Liouville theory,''
  JHEP {\bf 0506}, 014 (2005).

	
	
\end{thebibliography}
\end{document}